# Spin-torque nano-oscillator based on two in-plane magnetized synthetic ferrimagnets

E. Monteblanco[1,a)], F. Garcia-Sanchez[1], M. Romera[1,2], D. Gusakova[1], L. D. Buda-Prejbeanu[1], U. Ebels[1]

[1]*Univ. Grenoble Alpes, CEA, CNRS, Grenoble INP\*, INAC, SPINTEC, F-38000 Grenoble, France*
[2]*GFMC, Departamento de Física de Materiales, Universidad Complutense, Madrid, Spain.*
*\* Institute of Engineering Univ. Grenoble Alpes*
a) Electronic email: nmonteblanco@gmail.com

We report the dynamic characterization of the spin-torque-driven in-plane precession modes of a spin-torque nano-oscillator based on two different synthetic ferrimagnets: a pinned one characterized by a strong RKKY interaction which is exchange coupled to an antiferromagnetic layer; and a second one, non-pinned characterized by weak RKKY coupling. The microwave properties associated with the steady-state precession of both SyFs are characterized by high spectral purity and power spectral density. However, frequency dispersion diagrams of the damped and spin transfer torque modes reveal drastically different dynamical behavior and microwave emission properties in both SyFs. In particular, the weak coupling between the magnetic layers of the non-pinned SyF raises discontinuous dispersion diagrams suggesting a strong influence of mode crossing. An interpretation of the different dynamical features observed in the damped and spin torque modes of both SyF systems was obtained by solving simultaneously, in a macrospin approach, a linearized version of the Landau-Lifshitz-Gilbert equation including the spin transfer torque term.

## I. INTRODUCTION

The exceptional and multi-functional properties of spin-torque[1-2] nano-oscillators (STOs) made them promising candidates for a wide range of emerging technologies which span from microwave emitters[3] and detectors[4-5] to neuromorphic computing systems[6-10]. These devices use the transfer of angular momentum from a spin-polarized current to the local magnetization of a ferromagnetic layer[11,12] to induce self-sustained oscillations of the magnetization, which translate into a microwave signal whose frequency can be finely tuned with the applied direct current[13-17]. A widely studied structure of the STO is of the type (AF/SyF/MgO/SL) with in-plane magnetization, including a synthetic ferrimagnet structure (called SyF-Polarizer) pinned via exchange bias to an antiferromagnetic layer (AF) and a single ferromagnetic layer (SL)[18-20]. A standard SyF layer is composed of two ferromagnetic layers, a top (TL) and bottom (BL) layer coupled antiferromagnetically through a thin metallic layer via the Ruderman-Kittel-Kasuya-Yosida (RKKY) interaction.[21,22] In STOs, spin-polarized electrons affect damped oscillations by modifying damping and thus linewidth and amplitudes and overcoming a critical current density ($j_c$) polarized electrons can induce the steady-state oscillations (STT excitations). Steady-state oscillations can be obtained in both parts, SyF or SL, by changing the polarity of the current. The STT excitations of a SyF structure present some advantages in comparison with the SL excitations, as the higher spectral purity (smaller linewidth), zero field excitations[23,24], frequency tuning as a function of the current with the possibility to change from redshift ($df/dj_{app}<0$) to blueshift ($df/dj_{app}>0$) applying an in-plane field and achieving large thermal stability[25-29]. However, since the SyF is pinned by an antiferromagnet, achieving steady-state excitations requires a relatively large critical current. The use of exchange-coupled layers with perpendicular anisotropy[30,31] or in-plane magnetized magnetic layers[32-34] have been shown to increase the magnetic stiffness and can potentially improve the performance of the oscillator. Replacing the standard free layer SL by an unpinned SyF layer is thus of potential interest towards improving the microwave properties of spin torque oscillators.

In this article, we report the main static and dynamic features of a spin torque oscillator based on two SyFs with the following structure: IrMn(6.1)/SyF-Polarizer/MgO(0.9)/SyF-FL, where numbers represent thickness in nanometers. The composition of the SyF-Polarizer and SyF-FL are CoFe(1.8)/Ru(0.4)/CoFeB(2) and CoFe(0.5)/CoFeB(3.4)/Ru/CoFe(3.6) respectively. The structure of this manuscript is the following: section II introduces the numerical techniques used to predict some features of the STO dispersion diagrams and it is devoted to the macrospin analysis of the STO structure, section III presents the experimental characterization of the STO dynamics. Section IV presents the discussion and conclusions. In the following, the nano-oscillator based on a double SyF will be called D-SyF for simplicity.

## II. NUMERICAL ANALYSIS

Two types of numerical studies were performed in the framework of the macrospin approximation: *(1) c*omputation of the hysteresis loops (MH) and magnetoresistance loops (MR) using the minimization of the energy and *(2)* stability analysis based on the linearization of the Landau-Lifshitz-Gilbert (LLG) equation enhanced with the spin transfer torque term (Slonczewski term) around the equilibrium positions, in order to find instabilities due to the applied current and field. The D-SyF under study presents the following structure: AF/SyF-Polarizer/insulator barrier/SyF-Free layer, where the bottom layer of the SyF-Polarizer is pinned into the positive x direction by an exchange bias field ($H_{eb}$) induced by an antiferromagnetic layer, see Figure 1.

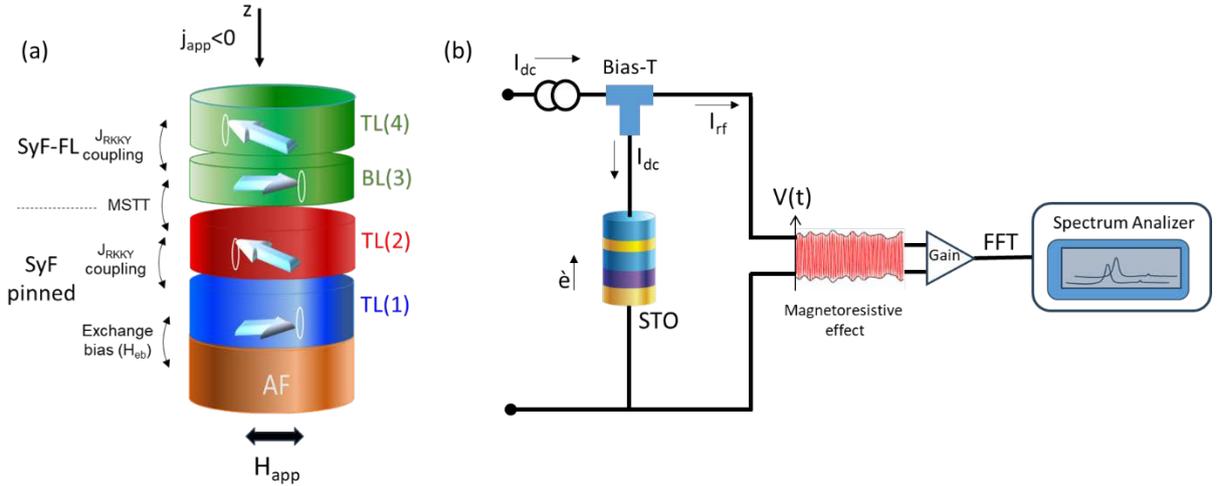

Fig. 1. (a) Schematics of the double SyF oscillator structure with the labels used in this article. Negative applied current corresponds to electrons flowing from SyF-FL to SyF-Polarizer. (b) Schematic of the standard RF setup.

### II.1. Simulation of static hysteresis loops

In order to understand the complex frequency dispersion diagrams of the D-SyF oscillator, its hysteresis loop has been simulated as the first step considering the different coupling between the ferromagnetic layers. The total free energy density of the coupled system is written as $\sigma_{tot} = \sum_i \sigma_{tot,i} = \sum_i(\sigma_{int,i} + \sigma_{ext,i})$ where the internal and external component for each layer are defined as follow,

$$\sigma_{int,i} = \sigma_{zeem,i} + \sigma_{an,i} + \sigma_{d,i} + \sigma_{eb,i} \quad (1)$$
$$\sigma_{ext,i} = \sigma_{RKKY,i} + \sigma_{dip,ij} \quad (2)$$

The magnetic layers are labelled as *i, j*=1 to 4 and $i \neq j$, corresponding to the structure on Figure 1. The model includes the different internal energies contributions such as the demagnetizing $\sigma_d$ and magnetocrystalline anisotropies $\sigma_{an}$, exchange bias $\sigma_{eb}$ (only for the BL of the SyF pinned) and the Zeeman terms $\sigma_{zeem}$. Also, we include the Ruderman-Kittel-Kasuya-Yosida (RKKY) interlayer exchange interaction $\sigma_{RKKY}$ (internal to each SyF) and the dipolar stray field[35-36] $\sigma_{dip}$, see Eq. (2). The RKKY coupling is taken into account only between the layers 1-2 and 3-4 ($\sigma_{RKKY,1(3)}= \sigma_{RKKY,2(4)}$) and the dipolar field between the four layers ($\sigma_{dip,ij}= \sigma_{dip,ji}$). More details about each term of the equations are found in Ref. 34 and 38. Two different RKKY coupling constants were considered: $J_{RKKY}$ =-0.1mJ/m$^2$ and -1.5mJ/m$^2$ for the SyF-FL (weak coupling) and SyF-Polarizer (strong coupling) layer respectively. The influence of a high or weak RKKY coupling in the hysteresis loops (MH) and magnetoresistance curves (MR) has been shown in previous studies[34]. The structure considered is not compensated, i.e., the product ($M_S*t*S$) for the TL and BL of each SyF are not close, ($M_S*t*S$)$_{TL,SyF-FL}$= 35.12 µA nm$^2$, ($M_S*t*S$)$_{BL,SyF-FL}$=28.79 µA nm$^2$, ($M_S*t*S$)$_{TL,SyF-Polarizer}$=14.06 µA nm$^2$, and ($M_S*t*S$)$_{BL,SyF-Polarizer}$=17.56 µA nm$^2$. The net magnetic moment of the SyF-FL is 6.33 µA nm$^2$ and for the SyF-Polarizer is 3.5 µA nm$^2$ so we should consider the stray magnetic field (dipolar field) between both SyFs. This is fundamental to understanding the magnetization dynamics of this structure.

|  | SyF Pinned Layer | | SyF Free Layer | |
|---|---|---|---|---|
| **Parameters** | *1* | *2* | *3* | *4* |
| $M_S$(kA/m) | 1470 | 1060 | 1112.5 | 1470 |
| K (J/m$^3$) | 7957.75 | | | |
| t (nm) | 1.8 | 2.0 | 3.9 | 3.6 |
| $M_S*t*S$ (µA.nm$^2$) | 17.56 | 14.06 | 28.79 | 35.12 |
| $N_x$ | 0.024952 | 0.027061 | 0.044620 | 0.042085 |
| $N_y$ | 0.046343 | 0.050280 | 0.083119 | 0.078372 |
| $N_z$ | 0.928705 | 0.922658 | 0.872260 | 0.879543 |
| **p** | - | **-m$_3$** | **m$_2$** | - |
| η | 0.3 | | | |
| α | 0.02 | | | |
| S (nm$^2$) | 6636.63 | | | |
| $H_{eb}$ (kA/m) | 79.5 | 0 | 0 | 0 |
| $J_{RKKY}$ (mJ/m$^2$) | -1.5 | | -0.1 | |

Table I. Parameters used in the numerical simulations. Here $M_S$ is the saturation magnetization, $K_u$ is a uniaxial magneto crystalline anisotropy constant (//ox axis in the plane), t is the film thickness, S is the surface, $N_x$, $N_y$, and $N_z$ are demagnetization factors, α is the damping constant. $H_{eb}$ is the exchange bias field that acts on the BL of the SyF-Polarizer and η is the spin efficiency.

Numerical simulation of the hysteresis loop and the magnetoresistance of the D-SyF oscillator are shown in Figure 2(a). These curves provide information about the relative orientation of the magnetizations of the different layers, as a function of the applied field. The MR was calculated with the scalar product of the magnetizations adjacent to the insulator barrier i.e. layers 2 and 3. The parameters are listed in Table I. In our samples, the anisotropy axis corresponds to the longer axis of the ellipse, and it is parallel to the X-axis. Arrows on Figure 2 represent the SyF-FL and SyF-Polarizer magnetizations respectively following the color convention of the layers in Figure 1. For relatively large positive values of the applied field [100mT, 400mT] the magnetization of both layers of the SyF-FL are parallel to the external field while the magnetization of the TL of the SyF-pinned is pointing in opposite direction, corresponding to the antiparallel state in the magnetoresistance curve (Figure 2b). Upon sweeping the field from positive to negative values, a first magnetization switching is observed at $\mu_oH_{sw} \approx$ +8.4 mT ($H^{sat+}_{SyF-FL}$ in Figure 2(a)). It corresponds to the switching of the magnetization of the BL of the SyF-FL, which presents a lower net magnetic moment in comparison with the TL. The magnetization switching of the BL of the SyF-FL is accompanied by a change from the antiparallel (high resistance) to the parallel (low resistance) state in the magnetoresistance curve, see Figure 2(b). Upon sweeping further the applied field towards negative values, the magnetization of the TL layer of the SyF-FL is inverted at $H^{sat-}_{SyF-FL}$ (Figure 2(a)), which does not affect the magnetoresistance (Figure 2(b)). From now on, we refer to the region of around 100mT between the two characteristic saturation field values ($H^{sat-}_{SyF-FL}$ and $H^{sat+}_{SyF-FL}$) as "plateau". It corresponds to the P state in the RH curve. Sweeping the field in the opposite direction (from negative to positive values) the bottom layer of the SyF-FL is reversed first again leading to a similar scenario except because the plateau region is shifted towards positive field values and now corresponds to the antiparallel state (Figure 2(b)).

**II.2. Dynamics features, the LLG equation**

The magnetization dynamics of the D-SyF structure is described in a macrospin approach solving the generalized Landau-Lifshitz-Gilbert (LLG) equations enhanced by the spin torque term[34,38]. The equation for each ferromagnetic layer is written as follows,

$$\frac{d\boldsymbol{m}_i}{dt} = -\gamma_0(\boldsymbol{m}_i \times \boldsymbol{H}_i^{eff}) + \alpha_i \left(\boldsymbol{m}_i \times \frac{d\boldsymbol{m}_i}{dt}\right) + \left(\frac{d\boldsymbol{m}_i}{dt}\right)_{STT} \quad (3)$$

$$\left(\frac{d\boldsymbol{m}_i}{dt}\right)_{STT} = \gamma_0 j_{app} G(\eta) \boldsymbol{m}_i \times (\boldsymbol{m}_i \times \boldsymbol{p}_i) \quad (4)$$

where the layers are identified by the number $i$=1 to 4, see structure in Figure 1(a). The vector $\boldsymbol{m}_i = \boldsymbol{M}_i/M_{Si}$ is the normalized magnetization vector, $M_{Si}$ is the corresponding saturation magnetization, $\gamma_0$ is the gyromagnetic ratio, $\alpha_i$ is the Gilbert damping constant. All the parameters are listed in Table I. $\boldsymbol{H}^{eff}_i = \boldsymbol{H}^{int}_i + \boldsymbol{H}^{coupling}_i$ is the effective field, composed of the internal field $\boldsymbol{H}^{int}_i$ and the coupling field $\boldsymbol{H}^{coupling}_i$ of the $i^{th}$ layer. The effective field is derived from the energy term of equations (1) and (2) for each layer. The last term is the spin torque term that affects the damping (second term in (3)). For this model we have not taken into account the effects of a mutual spin through the Ru spacer

During simulations, three types of dynamic couplings are considered: dynamic RKKY interaction between the two magnetic layers of each SyF, dynamic dipolar interaction between the four magnetic layers of the oscillator, and the mutual spin transfer torque (MSTT) only between the BL of the SyF-FL and the TL of the SyF-Polarizer, see Eq (4). Dynamic RKKY interaction and dynamic dipolar interaction are conservative couplings included in the precession term of the LLG equation (first term in Eq. (3)), without an energy loss. The Gilbert term and the MSTT, are dissipative couplings considered by the damping and the spin-transfer torque term (second and last term in Eq. (3)). Here $j_{app}$ is the applied current density, the pre-factor $G(\eta)$ is given by Eq. 2 in Ref. 38, with the spin polarization efficiency $\eta$ and the unit vector $\boldsymbol{p}_i$ represents the direction of the spin polarization vector of electrons. We are only considering the STT between layers 2 and 3 with the polarizer of layer 2 is being layer 3 and the opposite as shown in Table I. The equation (3) should be read as follows: the spin-polarized electrons from the $\boldsymbol{m}_j$ layer reach the layer $\boldsymbol{m}_i$ where the intrinsic damping starts to be counteracted ($j_{app}<j_c$) (only for the correct sign of current) due to the spin transfer torque. These damped excitations produced by the applied current are called the STT damped modes in this manuscript. There are four STT damped modes such as the result of the hybridization of the isolated SyFs modes, due to the dipolar coupling present between the layers in the structure. Overcoming a critical current value ($j_c$) the damping is completely counteracted, and the magnetization state is destabilized, leading to switching to another stable state or into steady state oscillations (STT excitations). All excitations involve all four layers of the oscillator.

In the case of positive applied current ($j_{app}>j_c>0$) the BL of the SyF-FL magnetization is driven by the spin transfer torque of spin-polarized current and this leads to excitations dominated by the SyF-FL, in which the SyF-Polarizer participate following the dynamics due to the dipolar coupling. For the case of the SyF-Polarizer dominant precession ($j_{app}<j_c<0$), it is the TL magnetization of the SyF-Polarizer which is controlled by the spin transfer torque. As this structure is coupled by the dipolar field, the four layers move together at the same frequency. In the absence of current ($j_{app}=0$) the spin torque term disappears, and we can simulate the excitation modes in the linear regime.

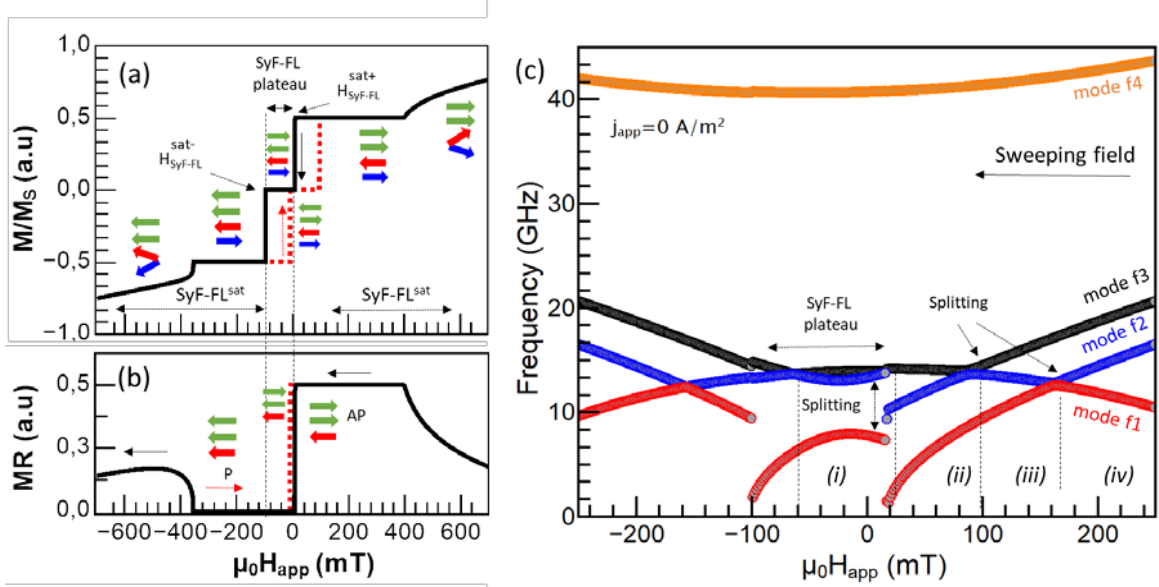

Fig. 2. (a)-(b) Hysteresis and magnetoresistance curves of the D-SyF oscillator. (c) Dispersion diagram of hybridized damped modes f1, f2, f3, and f4 for $j_{app}=0$. The magnetic field has been swept from positive to negative values. The regions *(i)-(iv)* are defined by the crossing of modes.

To calculate the STT damped modes ($j_{app} \neq 0$), the set of equations (3) are rewritten in the form of a matrix, giving a periodic solution for $\mathbf{m}_i$ and linearizing around an equilibrium position, $\mathbf{m}_i=(M_{eq},0,0)+(0,n_{y,i},n_{z,i})e^{pt}$. The corresponding equilibrium positions were extracted from the hysteresis loop. The eigenvalue of the characteristic matrix will be the complex number $p=\lambda+i\omega$, with a real part $\lambda$ corresponding to the decay rate of the excitation modes and the imaginary part $\omega$ corresponding to the $2\pi$ frequency of the excitations. The eigenvectors $(0,n_{y,i},n_{z,i})$ define the character of the oscillation mode and the amplitude of oscillation for each layer. Two types of instabilities can be described using the linearization. The first one corresponds to spin wave mode softening (f>0) and will coincide with the magnetization switching[32]. In the second type the real part of the characteristic eigenvalue can cross the zero axis. To monitor such instability, we have calculated the decay rates of the system as a function of the current density and the applied field $\lambda(j_{app},H_{app})$. This provides us with the main criteria to find the critical current density of the STT regime. A negative value means that the system relaxes into a stable state (damped regime), while a positive value indicates that the system becomes unstable, where one of the possibilities is the STT regime (auto oscillations). The passage from negative to positive defines $j_c$, i.e. the current where the decay rate is zero: $\lambda(j_{app},H_{app})=0 \rightarrow j_{app}=j_c$. This method does not consider temperature. Parameters used for the simulations were extracted from the experimental devices and correspond to SyF structures, which are not magnetically compensated.

The dipolar field was calculated numerically for both SyFs in the AP magnetic configuration ($H_{app}=0$), finding for the SyF-FL values around ±0.2 mT, for the BL and TL respectively. In the case of the SyF-Polarizer, the dipolar field calculated in the same X direction was -2.1 mT and -1.2 mT, for TL and BL respectively. Moreover, when the SyF-FL is already saturated (positive X direction) the dipolar field in the SyF-Polarizer increases up to -21.9 mT and -18.9mT. Therefore, due to the non-compensated SyF structures, we always expect the influence of the dynamical dipolar coupling in the magnetization dynamics.

Figure 2(c) shows the calculated STT damped mode frequencies at $j_{app}=0$ obtained by sweeping the applied field from positive to negative values. These modes are labeled f1, f2, f3, and f4 from low to high frequency. The dispersion diagram shows a well-defined mode splitting between modes 1 and 2, as well as two modes anti-crossings between modes 1 and 2, and 2 and 3 respectively. These effects are reminiscent of the splitting between the acoustic and optical-like modes provoked by the conservative RKKY interaction on single SyFs structures[37]. Here, the dipolar field between the four layers is responsible for the splitting and anti-crossings of the frequency dispersion diagram,[35] see Figure 2(c). Due to the crossing of modes, we define four regions, indicated in Figure 2(c) as *(i)-(iv)*. The region *(i)* is located for negative fields in the P state (low resistance state), and the three regions *(ii)-(iv)* on the AP state (high resistance state of the structure and interesting region to study the STT modes), see MR curve in Figure 2(b). In the following section, we study the evolution of the damped modes upon increasing the applied current and their stability.

### II.3. Magnetization stability analysis

We start the study of the stability of the magnetization precession around the equilibrium positions. Using the decay rate λ criteria, $\lambda(j_{app}, H_{app})<0 \rightarrow$ STT damped modes, $\lambda(j_{app}, H_{app})>0 \rightarrow$ STT modes and $\lambda(j_{app}, H_{app})=0 \rightarrow j_{app}=j_c$ introduced before, we can distinguish between stable and unstable regimes, by computing the critical current for each value of the applied field. Figure 3(a) shows the corresponding decay rates $\lambda(0, H_{app})$ of the damped modes f1, f2, f3, and f4 respectively. We will respect the color of modes defined in Figure 2(c). As it was expected for $j_{app}=0$, the decay rate of the four modes remain negatives, corresponding to the stability for the damped modes. Since the frequency dispersion vs. field was divided into several regions *(i)-(iv)*, due to the crossing of the damped modes, we notice that the decay rates suffer inflections or in some cases abrupt jumps from one of these regions to another one. It is important to remark that the mode f4 is less affected by the crossing of modes, showing the larger decay rate and without distortions in the whole range of applied field.

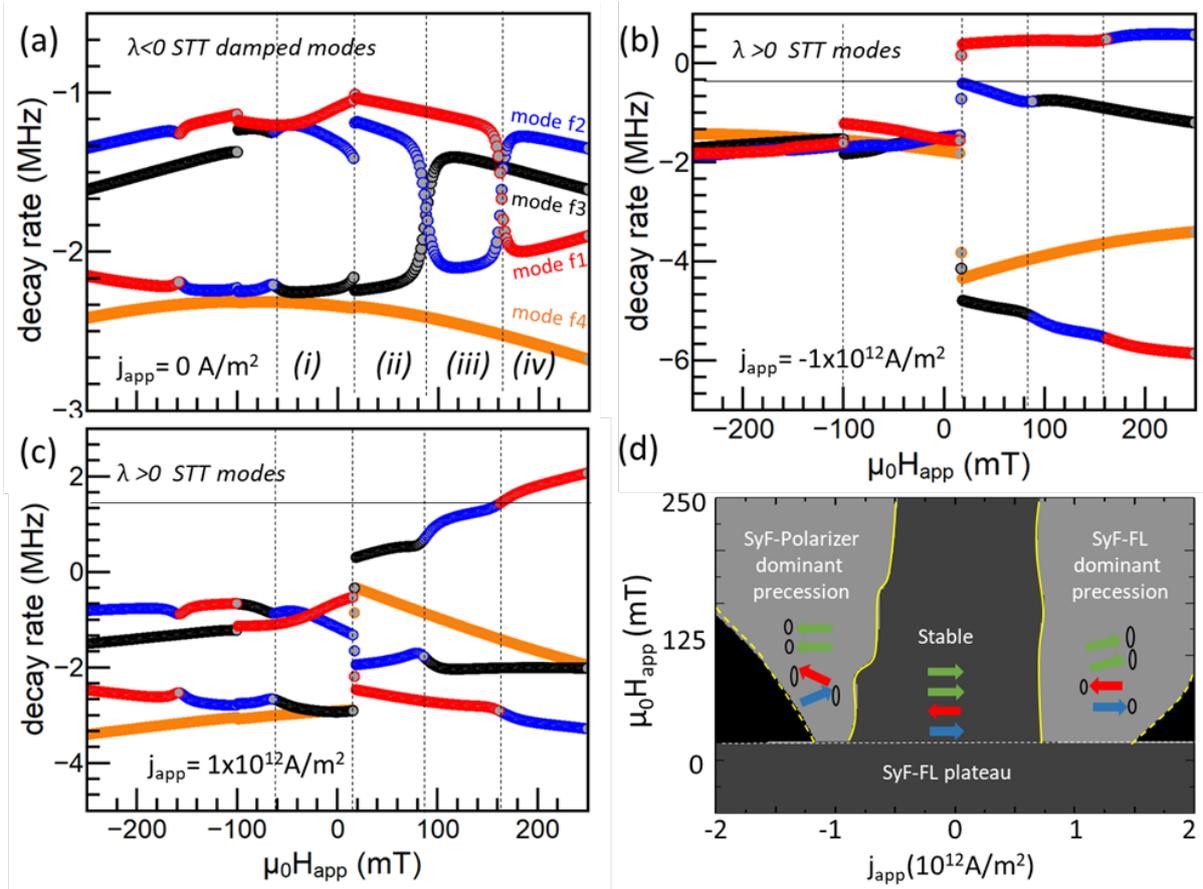

Fig. 3. Decay rate versus applied field dispersion diagrams, for the corresponding hybridized damped modes f1, f2, f3, and f4. Colors correspond to the modes defined in Figure 2(c). In (a) below the critical current $j_{app}=0$, in (b) for the SyF-FL dominant precession, $j_{app}=1 \times 10^{12}$ A/m$^2$ and in (c) for the SyF-Polarizer dominant precession, $j_{app}=-1 \times 10^{12}$ A/m$^2$. (d) State diagram $H_{app}$ vs $j_{app}$. Positive current corresponds to an electron flow from the TL of the SyF-Polarizer (red arrow) to the SyF-FL (green arrows). The light grey and black regions correspond to the unstable region of excitations, and the dark grey to the stable region. The black region corresponds to the switching of the SyF-FL.

When the positive current density is increased up to $j_{app}=1\times10^{12}$ A/m$^2$, the decay rate tendencies of the system change, see Figure 3(b). It is noticed that the decay rate for the modes f1 and f2 (dominated by the SyF-FL) are already positive $\lambda>0$, which is an indication that these modes evolve into the STT regime due to the damping compensation by the increase of the applied current. As this method is a linearization of the LLG equation, small magnetization precession, it is not possible to study the tendency of the decay rate when the system is already in the steady state regime, large magnetization precession. Applying negative current density, $j_{app}=-1\times10^{12}$ A/m$^2$, the system reaches the SyF-Polarizer dominant precession; see Figure 3(c). The decay rate is already positive for the modes f1, f2 and f3 (dominated by the SyF-Polarizer), generating STT modes in the frequency field diagram. For both senses of current density, it is observed a crossing of the decay rate in the *(i)* region. In conclusion, it is possible to obtain the critical current of the STT modes using the criteria $\lambda(j_{app},H_{app})=0$ (for positive and negative current density) and we can predict which of the modes will be stable or unstable. Sweeping the current density for each value of magnetic field we obtain the critical lines to build the state diagram $H_{app}$ vs $j_{app}$, shown in Figure 3(d).

The magnetic field was swept from positive to negative values. i.e. from the positive saturation magnetization of the SyF-FL until its plateau region in the P state (low resistance). The arrows correspond to the orientation of the magnetization of the four layers of the structure. The stable and unstable dynamical states are indicated by the dark and light grey regions, respectively. Excitations dominated by the SyF-FL (SyF-Polarizer) are observed in the region of positive (negative) current. Yellow lines indicate the critical current densities, $j_{c,SyF-FL}$ and $j_{c,SyF-Polarizer}$. The critical current values obtained and shown in Figure 3(d) are referential due to the simplification of the model and to the parameters used in performing simulations (exchange bias, RKKY coupling, thicknesses, $M_S$ etc). The critical currents should be taken as an approximation when compared to experimental results. The dashed yellow lines represent the border with the switching of the SyF-FL magnetizations, represented by the dark black region. In the SyF-FL dominant precession region, two small deflections can be observed, which corresponds to the splitting shown in Figure 1(c). The critical current for the SyF-FL in the saturated state is around $0.9\times10^{12}$ A/m$^2$.

This numerical simulation framework allows to predict the magnetization dynamic behavior of the coupled SyFs of the STO structure and the critical current as a function of the applied field. As we will verify in the next section, this framework provides useful information to understand the complex frequency dispersion diagrams studied experimentally, a fundamental issue in designing STO devices.

## III. EXPERIMENTAL SECTION

In this section, we present the static and dynamic features of the D-SyF nano-oscillator. Measurements were carried out using a standard microwave measurement setup.

### III.1. Static measurements

The experimental results are obtained for magnetic tunnel junctions with the following structure: IrMn(6.1)/SyF-Polarizer/MgO(0.9)/SyF-FL, where numbers represent thickness in nanometers. The composition of the SyF-Polarizer and SyF-FL are CoFe(1.8)/Ru(0.4)/CoFeB(2) and CoFe(0.5)/CoFeB(3.4)/Ru/CoFe(3.6) respectively. The thickness of Ru in SyF-FL was selected to achieve a weak and negative RKKY coupling. Samples were grown by sputter-deposition and patterned into elliptical nanopillars (130nm x 65nm) with and area of 6636.63nm$^2$. The uniaxial shape anisotropy stabilizes the magnetization in the direction of the longest axis. We have characterized tens of devices which we classified in two categories depending on whether the TMR is above or below 50% (High-TMR and Low-TMR devices respectively)[16]. The MgO barrier of the latter is considered to have inhomogeneities and/or pin holes either caused during deposition or electrical characterization. In this work, we mainly focus on HTMR devices. However, due to the slightly high critical current of STT excitations dominated by SyF-Polarizer, those cannot be achieved easily in HTMR devices applying voltages below 400 mV (MgO barrier degradation). Thus, LTMR devices characterized by a smaller resistance are used to characterize STT excitations dominated by the SyF-Polarizer. Figure 4(a) shows the MR curves of a High-TMR (HTMR, red curve) device and a Low-TMR (LTMR, black curve) device respectively. The MR curves are in good agreement with the numerical simulations (Figure 2(b)), which provide information about the magnetic orientations of the layers as a function of the applied field.

It has been found during the optimization of D-SyF structures, (not shown in this article) that the roughness in a multilayer structure gives less control of the thicknesses of the layers on top, thus the RKKY coupling of the SyF-FL becomes weak, reducing the size of the plateau. As it was shown in previous studies[34], the size of the plateau region of a SyF is directly related to the strength of its RKKY coupling and increased by the exchange bias field. The SyF-FL plateau is located between two characteristic fields, the $H^{sat-}_{SyF-FL}$ and $H^{sat+}_{SyF-FL}=H_{sw}$, however, it is not evident in the MR curve due to the weak RKKY coupling. The plateau size will be estimated in the next section measuring the STT damped modes. Both magnetizations of the SyF-Polarizer remain in its AP configuration (plateau region) for a quite large range of applied field (>500mT) thus we can consider this static configuration for all our measurements, until the spin flop field ($H^{sf-+}_{SyF-Polarizer}$).

### III.2. Dynamic Measurements

In this section, the dynamical features of the D-SyF oscillator device are presented. Section (A) shows the study of the STT damped modes on the plateau region of the SyF-FL. Section (B) corresponds to the study of the STT damped modes on the AP region (high resistance state). Section (C) is focused on the analysis of the STT modes for SyF-FL and SyF-Polarizer dominant precession. The same sign convention of the numerical simulations is considered: electrons flow from the SyF-Polarizer towards the SyF-FL for positive applied current, promoting STT excitations in the SyF-FL. It is worth noting that during the STT measurements, the STT damped modes are also excited.

### A. STT damped modes, positive current

First, the STT damped modes were measured for a HTMR device (TMR=60%) around zero applied field. The corresponding excitation frequency dispersion as a function of the applied field for positive applied current $I_{app}$=1 mA is shown in Figure 4(b). The power spectral density (PSD) of the STT damped modes is shown on a logarithmic scale. The magnetic field was swept from positive to negative values. We identify three different regimes in Figure 4(b) corresponding to the three different magnetic configurations of the SyF-FL (plateau *(i)* and both saturation regions) described by the numerical simulations (Figure 2). Regions *(i)* and *(ii)* in Figure 4(b) correspond to regions *(i)* and *(ii)* in Figure 2(c). Experimentally, we identified region *(i)* between -27 mT<$\mu_0 H_{app}$<10 mT. Out of this range, the SyF-FL is saturated.

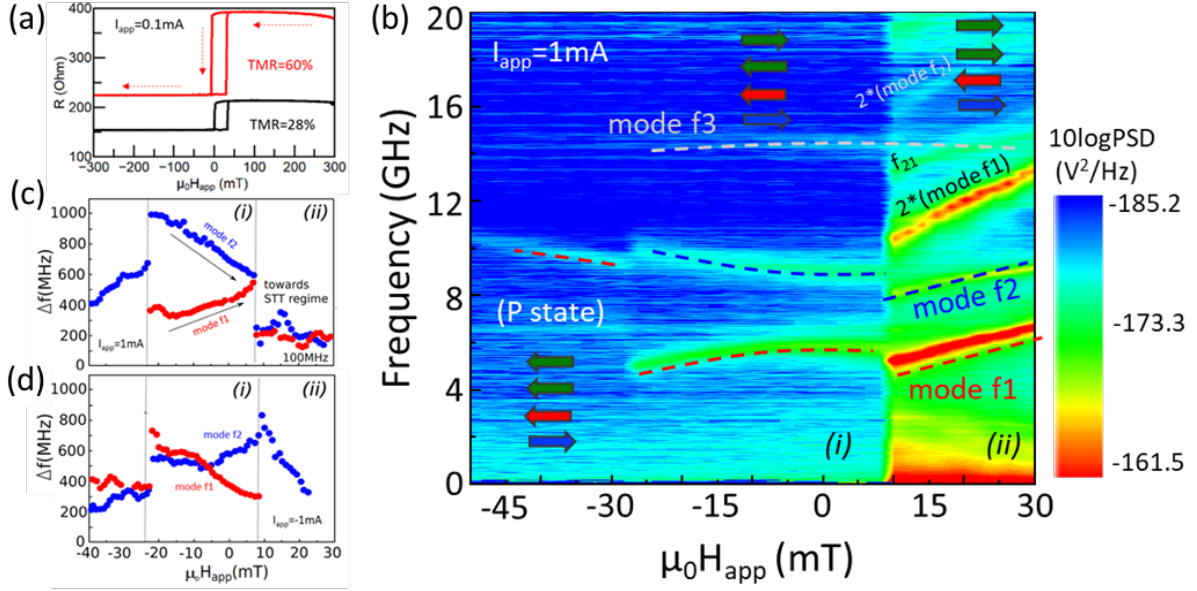

Fig. 4. (a) MR curves of an elliptical device (130x65nm$^2$) for HTMR device in red (TMR 60%) and LTMR in blue (TMR 28%) respectively. (b) Frequency vs. applied field of the dB (10 log of power spectral density (nV$^2$/Hz)) for positive applied current (I$_{app}$= 1 mA). The regions *(i)* and *(ii)* are identified in agreement with numerical simulations and the arrows correspond to the magnetic direction of magnetizations. The STT damped fundamental modes f1, f2 and f3 are identified, and the higher order damped modes f$_{11}$, f$_{21}$, and the harmonics 2f1, 2f2. The dashed lines are included to identify the modes only as a visual reference. Linewidth as a function of the applied field of the damped f1 and f2 STT damped modes for (c) positive and (d) negative applied currents (I$_{app}$ =±1 mA).

By comparing with the numerical simulations (Figure 2(c)), we can identify the STT damped modes f1, f2, f3 in regions *(i)* and *(ii)* in Figure 4(b), as well as other harmonics (2f1 is the 2nd harmonic of the f1 mode). In region *(i)* we observe the splitting and curvature of modes, which indicates weak conservative RKKY coupling of the SyF-FL, according to simulations. In region *(ii)* we observe other higher damped modes such as f$_{11}$, f$_{21}$, and the harmonics (2f1, 2f2), in agreement with results given in Ref. 39, where the finite size of the device generates quantized spin waves. As we will see in the next section, the appearance of higher-order modes has negative consequences on the microwave properties of the D-SyF oscillator.

As can be seen in Figure 4(b), the intensity of the different modes changes with the applied field (and applied current). As previously discussed, the amplitude of the magnetization precession and the linewidth should be proportional to the absolute value of the decay rate λ(j$_{app}$,H$_{app}$) of the STT damped modes in the linear regime. The tendency of the linewidth observed experimentally (Figures 4(c) and 4(d)) agrees well with the evolution of modes 1 and 2 obtained by numerical simulations in the region *(i)* (Figures 3(b) and 3(c)), indicating a good correspondence between our model and experiments. Large values of the linewidth (≈400-800 MHz), for a quality factor Q=Δf/f≈50-100, corresponding to STT damped modes (linear regime) were measured.

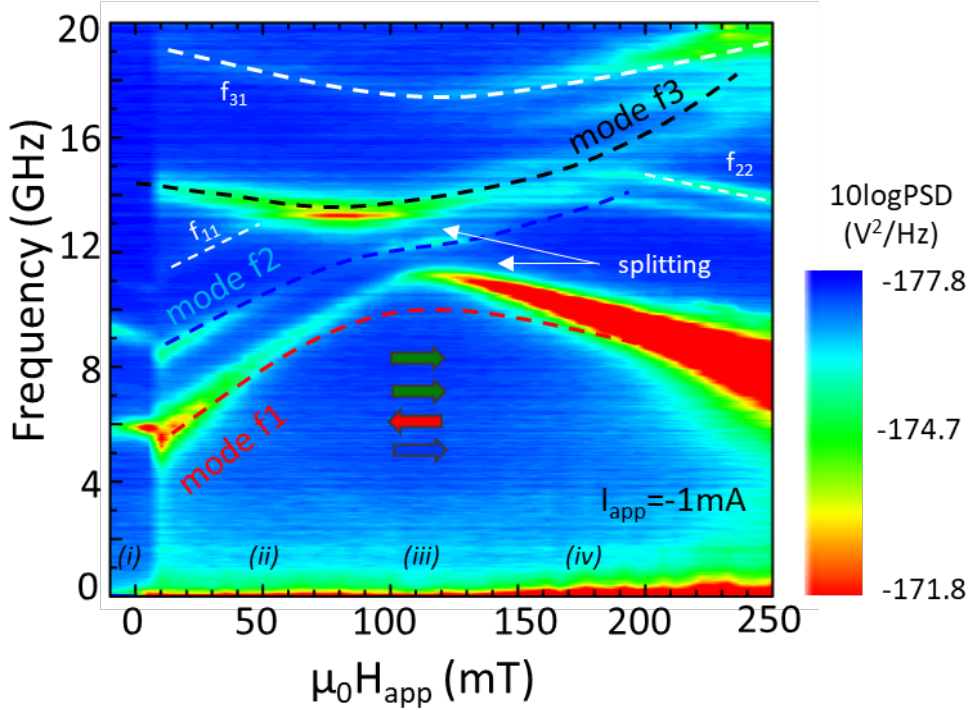

Fig. 5. Frequency vs. applied field diagram for regions (i)-(iv) for the device with TMR=60%, for $I_{app}$=-1 mA. Dashed lines correspond to the damped f1, f2, and f3 modes only as a visual reference. Arrows indicate the magnetization directions on the magnetic layers for the different regions.

### B. STT damped modes negative current

STT damped modes in both SyFs structures were measured by applying -1mA and they are introduced in the frequency dispersion diagram in Figure 5. For applied magnetic fields >10mT the SyF-FL is completely saturated (green arrows) while the SyF pinned remains in the AP magnetic configuration. By comparing with numerical simulations in Figure 2(c), we can also identify the regions *(ii)-(iv)* and modes f1, f2, and f3. The experimental STT damped modes follow the frequency-field dependence of the simulated modes, and they are accompanied by additional higher-order modes[27] ($f_{11}$, $f_{31}$, $f_{22}$). At low positive fields, modes f1 and f2 are separated 2.4GHz, as a result of RKKY and dipolar interactions within the SyF-FL (conservative couplings). Two additional mode splits are observed around 100mT, between modes f1 and f2 and between modes f2 and f3 respectively, in agreement with the numerical analysis, see region *(iii)* in Figure 2(c). In the next section, we discuss the evolution of the different STT damped modes into STT excitations in the SyF-FL or the SyF-Polarizer upon increasing the applied current for positive and negative values respectively.

### C. STT excitation modes

### C.1 HTMR device : SyF-FL dominantes STT mode

Steady-state oscillations (STT modes) were first analyzed by applying positive current, which favors STT modes dominated by the SyF-FL. Figure 6(a) shows the frequency-field dispersion curve with an applied current of $I_{app}$=0.92 mA. The observed SyF-FL dominated STT modes correspond to the evolution of the STT damped modes f1 and f2 on region *(ii)* and *(iii)*, as predicted by the stability analysis (Figure 3). The fundamental STT mode f1 shows several discontinuities around 40, 58 and 82 mT, in agreement with the macro-spin simulations (Figure 2(c)). These discontinuities are interpreted to be due to interactions between the steady state mode and other damped modes of the system through higher harmonics[29-35]. Indeed, modes $f_3$, $f_{11}$ and $f_{31}$ cross the second harmonic of the STT mode f1 (lines in Figure 5 and 6(a)). Non-linear mode interactions through higher harmonics can produce discontinuities and kinks in the fundamental STT mode f1[40-41].

Figure 6(b) displays the linewidth of the STT mode f1. The linewidth of STT mode 1 increases each time there is a discontinuity in the f-H dispersion, as expected from the interaction with other damped modes via higher harmonics[26]. The STT mode f2 is characterized by a much lower linewidth, reaching a minimum of 42 MHz, Q≈4.94, around 100 mT. At low fields (region *(i)*), the STT damped mode in the SyF-FL plateau shows a continuous frequency-field dispersion, without jumps or kinks, since there are no mode crossings in this range of field. Exciting STT modes in this region would potentially offer excitations at zero fields without linewidth enhancements due to undesired mode interactions. Due to the switching of the SyF-FL for an applied current below the critical current, it was not possible to obtain STT in the region *(i)*.

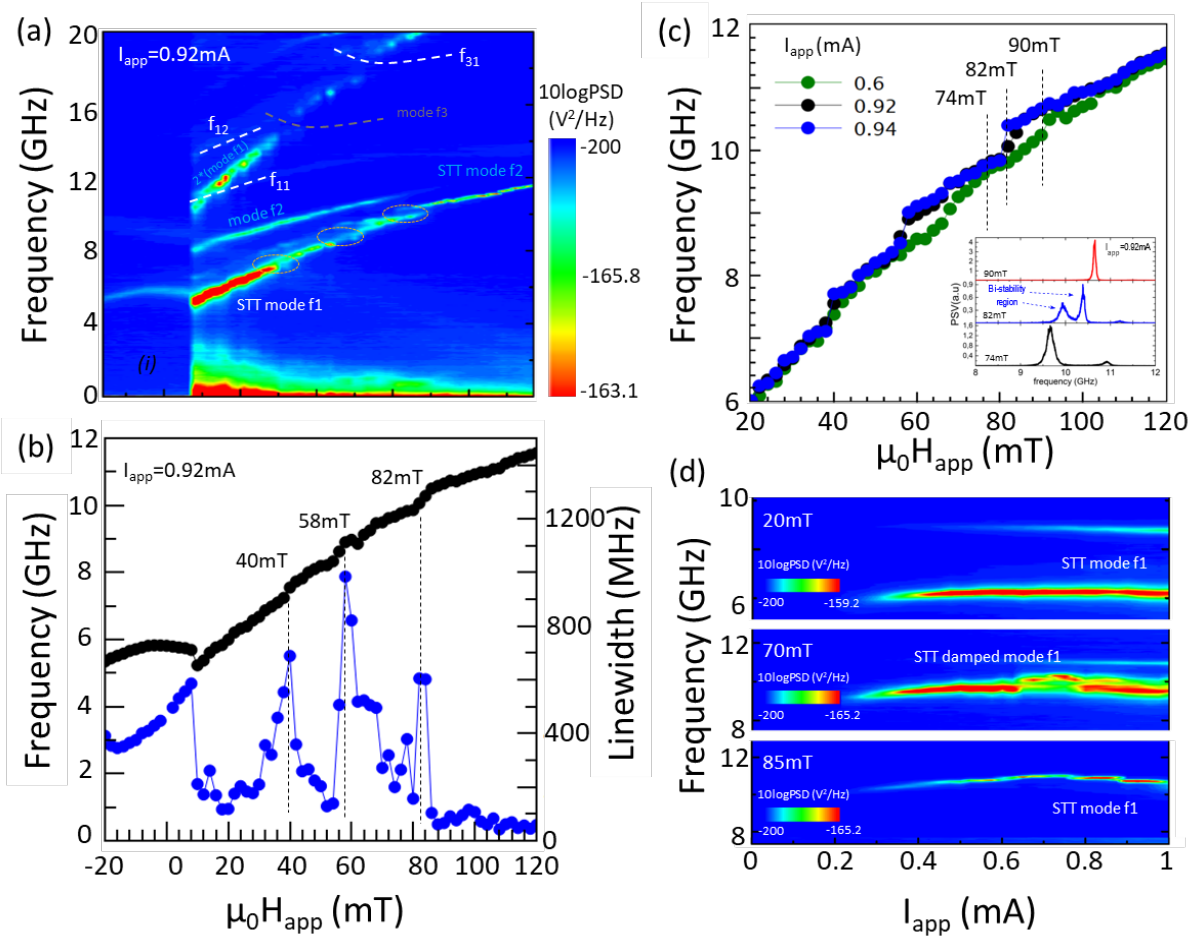

Fig. 6. (a) Frequency dispersion as a function of the field for the SyF-FL dominant precession, $I_{app}$=0.92 mA. The circles indicate the kink and jumps in the STT f1 mode. (b) Frequency (black) and linewidth (blue) versus applied field. Deviations in frequency correspond to an abrupt increase of Δf. (c) Frequency dispersion as a function of the applied field for the SyF-FL dominant precession at 0.6, 0.92 and 0.94 mA. Inset: peaks of modes interactions around the bi-stable region (around 82 mT). (d) Frequency dispersion as a function of the applied current for three values of applied field. Device HTMR (TMR=60%).

To study the discontinuities of the STT modes, the frequency field dispersion is plotted for three different values of the applied current (Figure 6(c)). The evolution from a continuous STT mode at 0.6mA (black curve), into a discontinuous STT mode with jumps and kinks at 0.94mA (blue curve). The PSD spectra at different fields around 82 mT are shown in the inset. The PSD spectra shows a region of bi-stability where two modes co-exist. This behavior has been reported to come from the interaction between a STT mode with other damped modes of the system through higher harmonics[37,39,41,42], where the apparent mode co-existence is indeed thermally activated mode hopping. This kind of discontinuities and jumps are observed only for large values of current, which implies a large amplitude of the magnetization precession and thus large dipolar interaction.

An interesting feature of the magnetization dynamic on SyF pinned structures is the possibility to tune the frequency-current dependence from redshift ($df/dj_{app}<0$) to blueshift ($df/dj_{app}>0$) by applying an in-plane field. Figure 6(d) shows the frequency as a function of the applied current for $\mu_0H_{app}$=20, 70 and 85 mT respectively. A transition from redshift to blueshift is observed upon increasing field from 20 to 85 mT. Interestingly both curves are continuous with no abrupt discontinuities. However, at $\mu_0H_{app}$= 70 mT, a bi-stable region characterized by mode is observed around $I_{app}$=0.82 mA (see dashed line).

### C.2. LTMR device

Achieving SyF-Polarizer dominant precession requires increasing the applied current above the breakdown voltage in HTMR devices. Thus, an LTMR device (TMR=28%) is used to pursue the study of the STT excitations dominated by the SyF-Polarizer. Figure 7 displays the STT modes of a LTMR device for positive (Figure 7(a)) and negative (Figure 7(b)) currents (±1.4 mA). The SyF-FL dominant excitations (Figure 7(a)) exhibit several discontinuities and kinks in the STT modes f1 and f2 modes and higher harmonics (2f1). Indeed, more discontinuities than the equivalent frequency-field dispersion of a HTMR device (Figure 6(a)) are observed. This is expected since the existence of pinholes in the tunnel barrier increases interlayer interaction. The frequency-field dependence becomes flat in the field region around 100mT because of the interaction between higher harmonics $f_{31}$ and 2f1. For negative current, the SyF-PL dominant precession shows a gap between the STT f3 and STT f1 modes (Figure 7(b)).

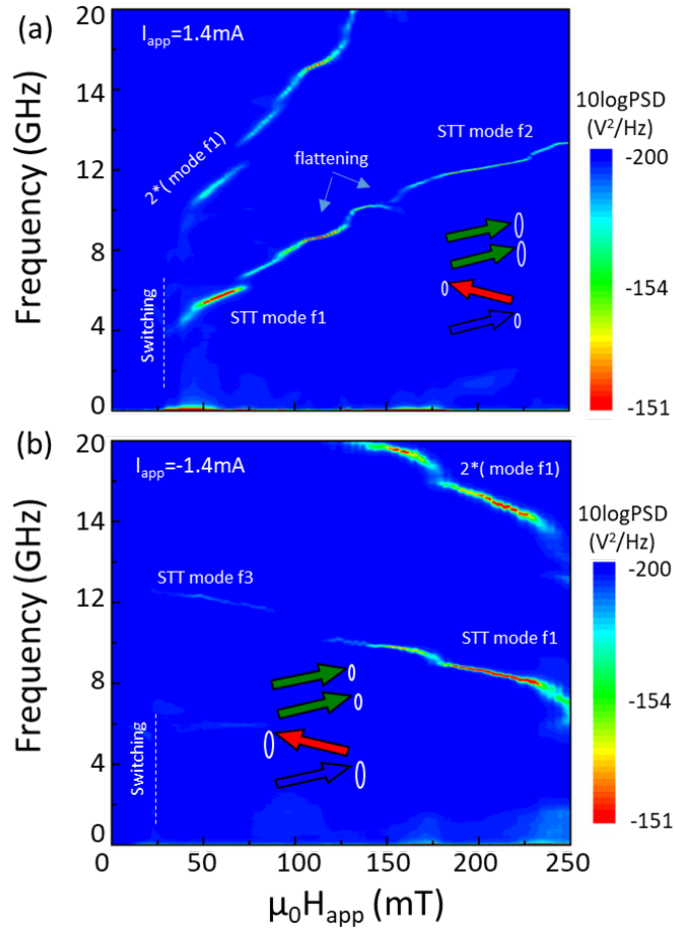

Fig. 7. (a) SyF-FL dominant precession, $I_{app}$=+1.4 mA and (b) SyF-PL dominant precession, for $I_{app}$=-1.4 mA. Interactions between the harmonics of the STT f1 and f2 modes with the damped or higher order modes will be transmitted as a form of kinks, deviations or jumps of the normal tendency of the frequency and linewidth versus applied field dispersion, generating discontinuities in the linewidth. Device with LTMR (TMR=28%).

The general tendency of the linewidth not shown is to decrease upon increasing the applied field, although several discontinuities (regions of linewidth enhancement) are observed corresponding to the interaction of the STT f1, f2, and f3 modes with damped and higher order modes, as in the HTMR device (Figure 6(b)). The local minimum linewidths measured for the SyF-FL dominant precession at 1.4 mA, were 20.5 MHz (Q≈1.73) at 112 mT, 13.6 MHz (Q≈1.1) at 194 mT and 12.9 MHz (Q≈1.17) at 214 mT. In the case of the SyF-Polarizer dominant precession at -1.4 mA, the minimum linewidths were 88MHz (Q≈8) at 46 mT, 53 MHz (Q≈5.57) at 150 mT and 46 MHz (Q≈5.41) at 214 mT. Interestingly, the minimum values of the linewidth associated to the SyF-FL dominant precession are smaller than the minimum linewidth associated to the precession of a single free layer on standard in-plane magnetized spin torque oscillators[22]. For the case of the SyF-Polarizer, the minimal linewidth was around ≈46 MHz. We note that continuous excitation branches can be found for specific conditions in the case of the SyF-Polarizer.

Finally, we analyze the redshift and blueshift regimes associated to SyF-FL dominant excitations. We note that the redshift and blueshift regimes have been studied only on SyF pinned structures[23,25] so far. Figure 8 displays the frequency-current dependence at different magnetic fields for the SyF-FL dominant regime. In (a)-(b) (20 mT-plateau region and 54 mT-P state respectively), the applied current remains below the critical current, therefore STT damped mode shown in (a) the standard redshift behavior with large linewidth Δf≈395-550 MHz (Q≈79-110). In (b) the STT damped mode shows the blueshift regime with a reduction of the linewidth by the current from Δf≈480 MHz (Q≈80) for low current for $I_{app}$ <1 mA to 125 MHz (Q≈17.8) for $I_{app}$>1.2 mA. Upon increasing the field, the critical current gets smaller. Thus, at $\mu_0 H_{app}$=70.7 mT in (c) the system evolves into the STT excitations. The frequency dispersion shows discontinuities and a transition from a blueshift regime to a redshift regime upon increasing current, around 1.1 mA. Looking at Figures 7 and 5, we observe that this discontinuity is a product of the crossing of the STT mode f1 with the damped mode f3, however, this transition is opposite to the previous observed (redshift into blueshift), and due to the complexity of these coupled systems. The mode transition into the redshift regime is accompanied by a reduction of the linewidth into Δf≈75 MHz (Q≈10) around 1.25 mA. At large fields (Figure 8(d)) a redshift to blueshift transition is observed upon increasing current, and a very low linewidth of Δf≈28 MHz (Q≈2.54) is obtained in the blueshift regime. This region of STT excitations at large fields and applied current in LTMR devices is of potential interest for applications as it offers the possibility of selecting excitations in the redshift or in the blueshift regime.

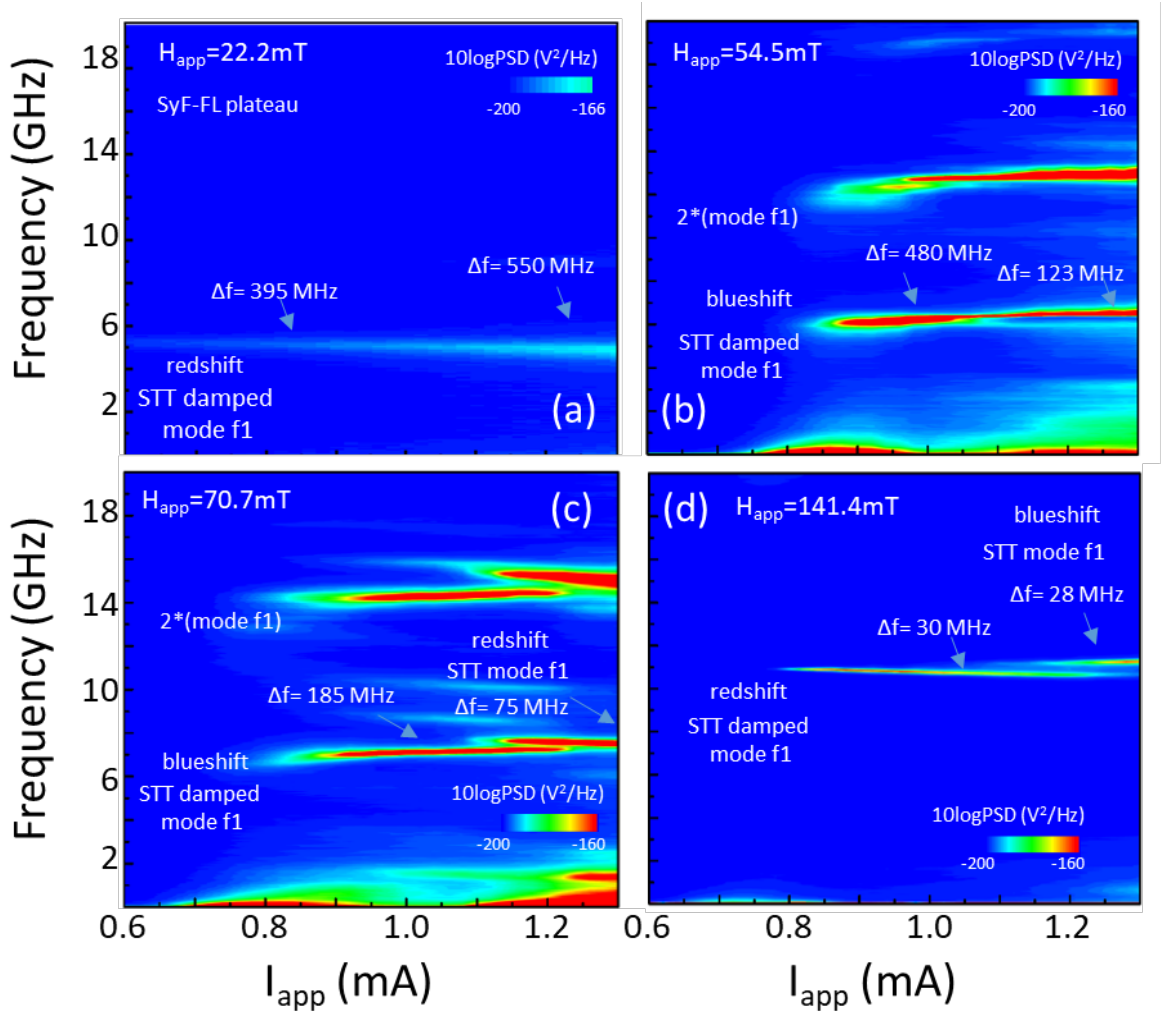

Fig. 8. Frequency as a function of the current density for $I_{app}>0$ for different field values (SyF-FL dominant precession). (a) STT damped mode f1 on the SyF-FL plateau region. (b) In the P state close to the switching field. The system shows a blueshift regime and large linewidth. In (c) the system shows a blueshift regime until $I_{app}=1.1$mA. Overcoming this value of current, the regime changes towards a redshift regime. In (d) the STT f1 mode shows first a redshift regime and after the splitting a blueshift regime with a minimum local linewidth of 28 MHz at 1.25 mA.

## IV. CONCLUSIONS

In this manuscript, we conducted a comprehensive investigation into the spin-transfer torque damped modes and steady-state oscillations of a spintronic nano-oscillator employing two SyF structures. The study involved both numerical simulations and experimental analyses. Numerical simulations were carried out using two different values of RKKY coupling, $J_{RKKY} \approx -0.1$mJ/m$^2$ and $J_{RKKY} \approx -1.5$ mJ/m$^2$ for the SyF-FL and SyF-Polarizer respectively. The small RKKY coupling of a SyF-FL eliminates the spin flop region, producing an abrupt switching of the magnetization layers and introducing a small plateau region ($\approx 40$ mT), in comparison with the corresponding of the pinned SyF-Polarizer (>600 mT). Static and dynamical experimental measurements confirmed the weak RKKY coupling of the SyF-FL on top of the structure. The experimental STT damped hybridized modes have been identified using numerical modeling. We found that it is possible to find the tendency of the linewidth and the PSD of the STT damped hybridized modes following the tendency of the decay rate ($\lambda$), also we can obtain the critical current of the STT modes. This study shows that the analysis of the stability predicts the state diagram of different structures, with an arbitrary number of layers. The frequency versus field dispersion diagrams for both devices (TMR 60% and 28%) show several discontinuities, attributed to the crossing of the STT mode or its harmonics with other damped hybridized or higher order modes, as was reported in Ref. 24, 36, 37 and 39. The introduction of a SyF-FL instead of a single layer in the STO based will generate more damped modes which produce more discontinuities in the STT modes. This last fact is also attributed to the dipolar coupling between the magnetic layers. The frequency current tuning shows two regimes for the SyF FL, the redshift ($df/dI_{app}<0$) and a blueshift ($df/dI_{app}>0$) which can be selected for different values of applied field. In the case of the SyF-Polarizer STT modes, a flat region ($df/dI_{app}\approx 0$) with a small redshift and blueshift regimes was found. We found the minimum linewidth on a LTMR device, around ≈12.9 MHz (Q≈1.17) for SyF-FL and around ≈46 MHz (Q≈5.41) for the SyF-Polarizer dominant precession, evidence of the stability of the system.

## V. ACKNOWLEDGE

M.R. acknowledges financial support from Spanish MIC, AEI and FEDER through Grant No. PID2020-116181RB-C33 (MCI/AEI/FEDER, UE) and from Comunidad de Madrid (Atracción de Talento Grant No. 2018-T1/IND-11935).